\documentclass[technote,onecolumn]{IEEEtran}
\makeatletter
\def\ps@headings{%
\def\@oddhead{\mbox{}\scriptsize\rightmark \hfil \thepage}%
\def\@evenhead{\scriptsize\thepage \hfil \leftmark\mbox{}}%
\def\@oddfoot{}%
\def\@evenfoot{}}
\makeatother
\makeatletter 
\newif\if@restonecol 
\makeatother

\usepackage[lined,boxed,linesnumbered]{algorithm2e}

\usepackage{colortbl, amsmath}
\usepackage{amssymb}
\usepackage{bm}

\usepackage{bbm}
\usepackage{subfigure}
\usepackage{cite}

\usepackage{tikz}
\usetikzlibrary{arrows,snakes,backgrounds}
\tikzstyle{router}=[circle,draw=black!50,fill=black!20,thick]
\tikzstyle{funcnode}=[rectangle,draw=black!50,fill=black!10,thick]
\tikzstyle{varnode}=[circle,draw=black!50,fill=black!20,thick]
\tikzstyle{na} = [overlay,ultra thick,<-]
\tikzstyle{nb} = [overlay,ultra thick,->]
\tikzstyle{nc} = [overlay,ultra thick,<->]

\definecolor{DarkBlue}{rgb}{0.1,0.1,0.5}
\definecolor{Red}{rgb}{0.9,0.0,0.1}
\definecolor{DarkGreen}{rgb}{0.1,0.5,0.1}

\newcommand{\CL}{\mathcal{L}}

\newcommand{\CP}{\mathcal{P}}

\title{Multi-path Probabilistic Available Bandwidth Estimation through Bayesian Active Learning}

\author{Frederic Thouin, Mark Coates, Michael Rabbat\\
Department of Electrical and Computer Engineering\\
       McGill University, Montreal, Canada\\
       frederic.thouin@mail.mcgill.ca, mark.coates@mcgill.ca, michael.rabbat@mcgill.ca}

\begin{document}

\tikzstyle{every picture}+=[remember picture]

\everymath{\displaystyle}

\maketitle

\begin{abstract}
Knowing the largest rate at which data can be sent on an end-to-end path such that the egress rate is equal to the ingress rate with high probability can be very practical when choosing transmission rates in video streaming or selecting peers in peer-to-peer applications.  We introduce \emph{probabilistic available bandwidth}, which is defined in terms of ingress rates and egress rates of traffic on a path, rather than in terms of capacity and utilization of the constituent links of the path like the standard available bandwidth metric.  In this paper, we describe a distributed algorithm, based on a probabilistic graphical model and Bayesian active learning, for simultaneously estimating the probabilistic available bandwidth of multiple paths through a network.  Our procedure exploits the fact that each packet train provides information not only about the path it traverses, but also about any path that shares a link with the monitored path. Simulations and PlanetLab experiments indicate that this process can dramatically reduce the number of probes required to generate accurate estimates.
\end{abstract}

\section{Introduction}

In many applications, especially peer-to-peer and overlay, it is beneficial to know what kind of performance guarantees end-to-end networks paths can offer.
More specifically, knowing the rate at which data can be sent between two nodes such that the injected traffic induces minimal congestion is very valuable to guide peer selection in peer-to-peer streaming, to choose transmission rates in video streaming or to choose client-server association in content delivery networks~\cite{jai:08}.
In this paper, we are interested in finding the largest rate at which data can be sent on an end-to-end path such that there is high probability that the output rate is almost the same as the ingress rate.  A network metric closely related to this value is available bandwidth, which is typically defined as the average unused capacity of a path over a given time interval.  

Estimating available bandwidth accurately involves saturating the path for a short period of time with a few high-rate packet trains. If this is done rarely, then the overhead is acceptable, but if many paths in a subnetwork are being monitored simultaneously, then due to shared links, the additional load becomes unacceptable and the measurement process itself can significantly bias the estimates.  We could sequentially apply the same strategy to all these paths independently. Not only is this approach time-consuming, but it is also inefficient in terms of resources, since we may probe the same link multiple times via different paths to obtain the same information.  

Intuitively, however, it should be possible to accurately and efficiently estimate the available bandwidth on multiple paths in a network by exploiting spatial correlation~\cite{hu:05,son:07}.  The available bandwidth of a path is limited by the \emph{tight link} --- the link with the least available bandwidth on this path.  Knowing something about the available bandwidth of one path reveals information about other paths that share some subset of links.  For example, if paths $P$ and $Q$ have different origins and the same destination, knowing that $P$ has less available bandwidth than $Q$ implies that the tight link of path $P$ must be on a link that is not shared by path $Q$.  This can help to localize the tight links to a particular segment, and consequently improve inference for other paths that overlap with the tight segment.

\subsection{Contributions}
\label{ssec:contrib}
Unlike previous work on available bandwidth estimation, which focuses on estimating the average unused capacity, we use a probabilistic notion of available bandwidth.  Namely, we focus on estimating the largest ingress rate for each path such that the destination will receive at nearly the same rate with high probability.  This is a natural way to quantify available bandwidth for streaming and file-transfer applications.  This paper proposes a method to simultaneously estimate the available bandwidth on multiple paths in the network.  Our goal is to accurately infer the available bandwidth on multiple paths without overwhelming the network with excessive measurement traffic.  Our approach to efficient probing is based on two underlying concepts.  First, we develop a probabilistic model that allows us to use the measurements taken on one path to draw inferences about the available bandwidth on other paths in the network.  This model accounts for variability in the measurements and also allows us to quantify how much uncertainty we have about estimates on each path.  Second, we design measurements in a sequential fashion using ideas from active learning.  After each measurement, we re-evaluate the uncertainty in our current available bandwidth estimates, and then select the next measurement (path and probing rate) so that it will maximize the gain in information.

To exploit correlations that exist between paths, we encode the overlap between different paths in a probabilistic graphical model (factor graph).  The graphical model contains one variable for each path, as well as one unmeasurable variable representing the available bandwidth on each link in the network.  The variables represent the available bandwidths on each path and link, respectively.  Correlation among different paths is encoded by connecting each path to the links it traverses through a function expressing the notion that the available bandwidth on the path is the smallest available bandwidth of all links.  Thus, the graphical model represents a joint probability distribution on available bandwidth on all paths and links in the network.  Our use of the graphical models formalism also allows us to exploit well-known efficient techniques for performing inference.  Given a collection of measurements, we can quickly form estimates of the marginal distributions of all path available bandwidths using belief propagation.

We select which measurement to collect next using ideas from active learning.  Similar to other bandwidth estimation techniques, such as Pathload~\cite{jai:03}, one measurement involves sending a quick burst of traffic along an end-to-end path at a fixed rate, and the measurement outcome indicates whether the probed rate is above or below the available bandwidth on the path.  At each measurement iteration, we select the path and rate at which to probe next in order to probabilistically bisect the uncertainty in our model, given the previous measurements.  This is analogous to performing a probabilistic binary search, and if there is just a single path being monitored, the approach is similar to related work for single-path bandwidth monitoring.  In our framework, the difference is that we determine both which path to probe and what rate to probe in an active, online fashion, based on the measurements we have already gathered over the entire network.  Simulations and PlanetLab experiments suggest that our framework significantly reduces the amount of probing traffic (by more than 50\%) in comparison to na\"ively probing sequentially on every path, without any loss in accuracy.

The rest of this paper is organized as follows.  In Sect.~\ref{sec:pab}, we introduce the notion of probabilistic available bandwidth.  In Sect.~\ref{sec:problem_statement}, we specify the estimation problem we are solving and state our assumptions.  In Sect.~\ref{sec:methodology}, we outline our proposed solution based on active learning, graphical models and belief propagation.  In Sect.~\ref{sec:exp}, we describe our measurement software implementation and show the results obtained from simulations and online experimentations.  In Sect.~\ref{ssec:related_work}, we discuss our contributions in relation to previous work in the area.  In Sect.~\ref{sec:conc}, we conclude and present future work.

\section{Probabilistic Available Bandwidth}
\label{sec:pab}

Our goal is to provide estimates of the {\em probabilistic available bandwidths} of multiple paths in a network. We use the term ``probabilistic available bandwidth'' to highlight the fact that the quantity we strive to estimate is not the same as the standard available bandwidth metric, which is defined in terms of the capacity and utilization of the constituent links of the path~\cite{jai:03}. As discussed in~\cite{liu:08}, there is a modelling assumption underpinning the available bandwidth estimation techniques that are based on the principle of self-induced congestion (sending packet trains and comparing ingress and egress rates)~\footnote{Neither Pathload~\cite{jai:03} nor pathChirp~\cite{rib:03}, two widely-cited bandwidth estimation schemes, are exactly based on rate comparisons. However, Pathload employs a test based on detection of an increasing trend in one-way delays, and this test is highly correlated with a rate comparison test. Although Jain et al. argue that these tests are not equivalent~\cite{jai:04}, it is relatively clear that any received packet train that passes the one-way delay test employed by Pathload (i.e., shows an increasing trend in delays), would have a measured egress rate substantially smaller than the rate of the injected probe. Pathload's one-way delay test is a more conservative test, but it detects essentially the same phenomenon as a test that compares ingress and egress rates. Similar comments apply regarding the identification of excursion regions and the resultant available bandwidth estimator employed by pathChirp~\cite{rib:03}.}.
The assumption is that the available bandwidth is equal to the largest probe rate at which the egress rate is (approximately) equal to the ingress rate, and that beyond this rate, the egress rate will drop below the ingress rate. This assumption is true for fluid traffic models and a single hop path. Liu et al. show that it is only approximate for multi-hop paths and other traffic models~\cite{liu:08}. In most cases, the egress rate will drop below the ingress rate at a probing rate well below the available bandwidth, leading to under-estimation.

Our metric of interest is the rate at which we can transmit data along a path so that there is a specified probability of inducing congestion, which occurs when the egress rate drops significantly below the ingress rate. We therefore specify the {\em probabilistic available bandwidth} metric directly in terms of ingress rates and egress rates of traffic on a path; we do not require a direct connection to path or link utilization.

We represent a network by a set of $N$ directed links $\CL$ and a set of $M$ paths $\CP$.  We assume that at the start of each link is a store-and-forward first-come first-served router/switch that dictates the behaviour of the link (in terms of delay, loss, utilization) and that the routing topology of this network is known, as embodied in the set of paths $\CP$, and that it remains fixed for the duration of our experiments. We denote by $r_p$ the ingress rate of a traffic flow on a path $p$ and by $r'_p$ the egress rate of the same traffic flow. For a link $\ell$, we adopt similar notation, denoting the ingress rate of the flow $r_{\ell}$ and the egress rate $r'_{\ell}$. We are interested in determining the largest ingress rate at which we can send a traffic flow along a path while achieving, with specified probability at least $1-\delta$, an egress rate that is almost as large as the ingress rate. More formally, for given $\epsilon > 0$ and $\delta > 0$, we seek the largest $r_p$ such that $\Pr(r'_p > r_p - \epsilon) > 1 - \delta$, where the probability is defined over all possible flows of rate $r_p$ that can complete transmission during a specified measurement period.  We denote the largest such ingress rate by $y_p(\epsilon,\delta)$ and refer to it as the $(\epsilon, \delta)$-available bandwidth (the probabilistic available bandwidth) for path $p$. For a link $\ell$ we use the notation $x_{\ell}(\epsilon,\delta)$.  Our inference procedure relies on a relationship between the $(\epsilon, \delta)$-available bandwidth of a path and the $(\epsilon, \delta)$-available bandwidths of its constituent links. We now develop this relationship and specify the assumptions underpinning it. We first note that $(\epsilon, \delta)$-available bandwidth is non-decreasing in $\epsilon$ or $\delta$.

For a path $p$ consisting of the set of links $L_p = \{1,2,\dots,n\}$, it is possible to identify small constants $0<\epsilon_{\ell} < \epsilon$ and $0<\delta_{\ell} <\delta$ such that:
\begin{equation}
\Pr(r'_{\ell} \leq r_{\ell} - \epsilon_{\ell}) \leq \delta_{\ell}\quad {\mathrm{ for\,\, all }} \quad r_\ell \leq y_p(\epsilon,\delta). \label{eq:linkcond}
\end{equation} 
but 
\begin{equation}
\Pr(r'_{\ell} \leq r_{\ell} - \epsilon_{\ell}) > \delta_{\ell}\quad {\mathrm{ for\,\, all }} \quad r_\ell > y_p(\epsilon,\delta). \label{eq:linkcond2}
\end{equation} 

This follows from the fact that the rate decrease on a path is always greater than the rate decrease on any of its constituent links. Suppose for the moment that the condition $r'_{\ell}>r_{\ell}-\epsilon_\ell$ holds for each link. Then we have the following relationship between the path and link ingress and egress rates:
\begin{eqnarray*}
r_1 &=& r_p\\
r_2 &=& r_1' > r_p - \epsilon_1 \\
r_3 &=& r_2' > r_p - \epsilon_1 - \epsilon_2 \\
&\vdots& \\
r'_p &=& r'_n > r_p - \sum_{i=1}^n \epsilon_i.
\end{eqnarray*}
We can apply the union bound on the links to establish:
\begin{equation*}
\Pr\left(\bigcup_{{\ell} \in L_p} \{r'_{\ell} \le r_{\ell} - \epsilon_{\ell}\}\right) \le \sum_{{\ell} \in L_p} \delta_{\ell}.
\end{equation*}
This relationship implies the following:

\begin{equation}
\Pr\left(r'_p > r_p - \sum_{{\ell} \in L_p} \epsilon_{\ell}\right) \geq 1 - \sum_{{\ell} \in L_p} \delta_l. \label{eq:pathcond}
\end{equation}

For sufficiently large $(\epsilon,\delta)$  we assume that it is possible to identify small constants $0<\epsilon_{\ell} < \epsilon$ and $0<\delta_{\ell} <\delta$ that satisfy (\ref{eq:linkcond}) for all $\ell \in L_p$, such that $\textstyle{\sum}_{{\ell} \in L_p} \epsilon_l < \epsilon$ and $\textstyle{\sum}_{{\ell} \in L_p} \delta_l < \delta$.  Moreover, we assume that there is a \emph{tight link} ${\ell}^* \in L_p$ which essentially determines the probabilistic available bandwidth on the path $p$. This means that it is possible, for all ${\ell}\in L_p$, ${\ell} \ne {\ell}^*$, to identify $\epsilon_{\ell}\ll \epsilon$ and $\delta_{\ell}\ll \delta$ that satisfy (\ref{eq:linkcond}). In the case of ${\ell}^*$, however, the smallest $\epsilon_{{\ell}^*} < \epsilon$ and $\delta_{{\ell}^*} <\delta$ pair that satisfy (\ref{eq:linkcond}) have the property $\epsilon_{{\ell}^*} \approx \epsilon$ and $\delta_{{\ell}^*} \approx \delta$.

The tight link assumption implies that $\textstyle{\sum}_{{\ell} \in L_p} \epsilon_l \approx \epsilon_{{\ell}^*} \approx \epsilon$ and $\textstyle{\sum}_{{\ell} \in L_p} \delta_l \approx \delta_{{\ell}^*} \approx \delta$. This property, together with (\ref{eq:linkcond}), ((\ref{eq:linkcond2}), and (\ref{eq:pathcond}), imply that 
$y_p(\epsilon, \delta) \approx x_{{\ell}^*}(\epsilon, \delta)$. Another way of interpreting this assumption, is that the $(\epsilon,\delta)$-bandwidth of any link $\ell \in L_p$, $\ell \neq {\ell}^*$ is significantly greater than $y_p$. In other words, for sufficiently large $(\epsilon,\delta)$, the $(\epsilon,\delta)$-available bandwidth is determined by the minimum $(\epsilon,\delta)$-available bandwidth of its constituent links.

\section{Problem Statement}
\label{sec:problem_statement}
Now that we have defined the $(\epsilon,\delta)$-available bandwidth of a path, we can formalize our problem statement. For a specified $(\epsilon,\delta)$ and network $(\CL,\CP)$, we employ packet train measurements, with the goal of forming estimates of the $(\epsilon,\delta)$-available bandwidths of all paths in the network. Let the available bandwidth of each link $\ell$ and path $p$ be modelled as discrete random variables $x_{\ell}$ and $y_p$ (e.g., $\Pr(y_p = r)$ being the probability that the $(\epsilon,\delta)$-available bandwidth on path $p$ is $r$).  For each packet train measurement, we evaluate a binary outcome $z$ which specifies whether the egress rate of the train was within $\epsilon$ of the ingress rate.  Then at any given instant $k$, we are interested in the marginal $\Pr(y_p | z_1,...z_k)$ for every path $p$, where $z_k$ is the outcome of the measurement at iteration $k$.

To obtain an estimate of that distribution, we adopt a Bayesian estimation framework.  We specify a likelihood function, learned from empirical training data, that relates this binary outcome to the difference between the probe rate and the underlying $(\epsilon,\delta)$-available bandwidth of the probed path.  Our non-informative prior model is a uniform distribution over the range $[B_{\min},B_{\max}]$ for the $(\epsilon,\delta)$-available bandwidth of each link, where $B_{\min}$ and $B_{\max}$ are estimates of the minimum and maximum available bandwidths of links. 

Our goal is to sequentially identify, using active learning, a sequence of measurements that allows us to form estimates of the $(\epsilon,\delta)$-available bandwidths, such that the credible intervals of the estimates are acceptably tight.  We use the confidence range $\beta$ and confidence level $\eta$ as stopping criteria:

\begin{eqnarray}
	\eta_p = \min_{ub_{p}-lb_{p}} \int_{lb_{p}}^{ub_{p}} \hat{p}(y_p | z_1,...,z_k) d{y_p} &\geq& \eta \label{eq:completion1}\\
	\beta_p = ub_{p}-lb_{p} &\leq& \beta \label{eq:completion2}
\end{eqnarray}

\noindent where $\eta_p$ is the confidence level (percent of the probability mass), $\beta_p$ is the confidence range of path $p$ with lower bound $lb_p$ and upper bound $ub_p$, and $\hat{p}(y_p)$ is the empirical pdf of $y_p$.  The procedure terminates when all paths satisfy these two conditions.  To ensure termination, we stop if we reach a specified maximum number of measurements whether the estimates meet the stopping criteria or not.The secondary objective is to sequentially choose these measurements (the path to probe and the probing rate), taking into account the past binary outcomes, such that the total number of measurements required to construct the estimates is minimized.

\subsection{Assumptions}

We assume the routing topology is known, i.e. that we are able to construct the network $(\CL,\CP)$.  
We first infer links and the mapping from IP addresses to routers using $\texttt{traceroute}$ (which has been known to inflate the number of observed routers, record incorrect links and bias router degree distributions~\cite{she:08}).  However, due to our problem formulation, our approach is robust to these kind of errors in the inference procedure; additional links can lead to an increase in the number of measurements, but has little impact on the accuracy of the estimates.  Ultimately, if our validation shows that our estimate of probabilistic available bandwidth is accurate, the accuracy of the actual topology is not relevant.

We use logical topologies (combine all links that are in a series) rather than routing topologies to reduce the number of links and the complexity of the factor graph.  We are able to do so without affecting the results since the available bandwidth of a two or more links in series will be limited by the link with the smallest available bandwidth.  Thus, we operate on the logical topology without loss of generality.

We assume that the routing topology is stable during the estimation procedure, which only takes up to a few minutes for moderate-size topologies (see Sect.~\ref{sec:exp}) - the scale over which the proposed algorithm is most applicable.  Throughout all our experiments (performed at different times of the day and  different days), the resulting logical topologies were stable.  

As explained in Sect.~\ref{sec:pab}, we assume the existence of a tight link on every path; the link with minimum $(\epsilon, \delta)$-available bandwidth on path $p$ that determines the value of $y_p$.

\section{Methodology}
\label{sec:methodology}

Our approach to network-wide available bandwidth estimation employs packet train measurements.  Each packet train probes one path at a specified rate.  When we probe path $p$ at a rate $r_p$, the outcome of the measurement is a binary variable $z=\mathbf{1}\{r'_p > r_p - \epsilon\}$, indicating whether the output rate $r_p'$ was within $\epsilon$ of the input rate.  Each measurement is then a $(p,r_p,z_{RDT})$ triple, where $z_{RDT}$ the binary outcome of the rate difference test (RDT).  The information obtained from all previous measurements is fused and summarized after each measurement iteration by maintaining an estimate of the available bandwidths for each path. The way information is shared amongst nodes in a factor graph is by using message passing, or belief propagation~\cite{fre:98}.  This procedure allows us to estimate the marginals in a factor graph and to update these distributions easily when we have new information from a measurement. Of course, we would like to obtain network-wide available bandwidth estimates using as few probes as possible.  To accomplish this, we adopt an active learning approach~\cite{elg:93}, sequentially choosing which path to probe next based on the measurements already obtained.  Our approach is summarized in Algorithm~\ref{alg:main}.

\begin{algorithm}
\caption{Multipath available bandwidth estimation algorithm}
\label{alg:main}

create factor graph using known topology\;
\While{\# measurements $<$ MAX MEAS}
	{
		choose path to probe next\;
		choose rate to probe\;
		take new measurement\;	
		update marginals using belief propagation\;
		\uIf{$\forall$ paths $p$: (\ref{eq:completion1}) and (\ref{eq:completion2}) are satisfied}{
			break;
		}
	}

\end{algorithm}

\subsection{Rate Difference Test}

Whenever $r_p > y_p$, one of the links of path $p$ receives more traffic than it can service, which causes queuing along the path; ultimately the rate observed at the receiving end, $r'_p$, should be lower than $r_p$. The special case where $\epsilon=0$ in the RDT is then equivalent to using the input/output rate ratio employed in several previously proposed available bandwidth estimation algorithms. We define the outcome of the rate difference test as $z_{RDT} = \mathbf{1}\{r'_p \geq r_p-\epsilon\}$.

\begin{figure}[!h]
	\centering
	\includegraphics[width=0.3\linewidth]{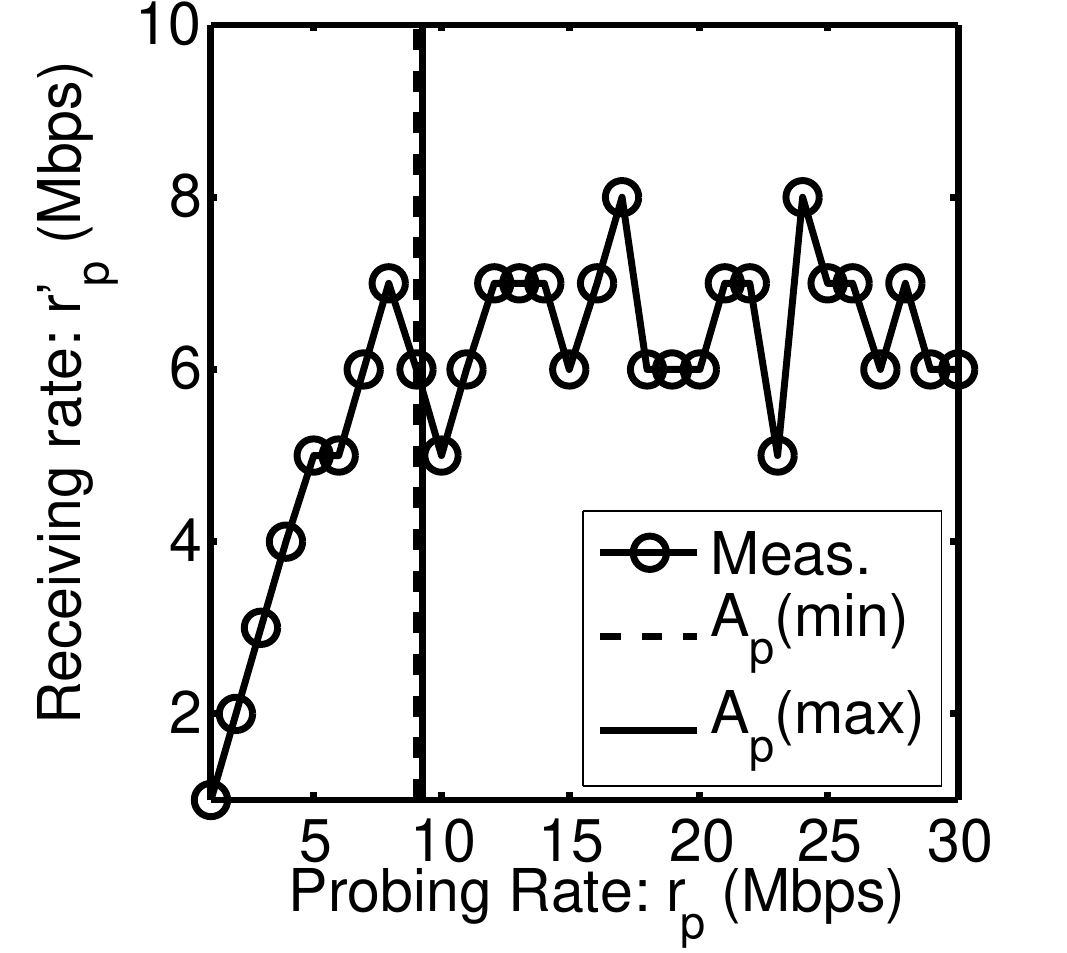}
	\caption{Observed receiving rate $r'_p$ as a function of the probing rate $r_p$.  As an indication, we show $A_p(min)$ and $A_p(max)$, the bounds of the interval where $y_p$ lies according to Pathload results.}
	\label{fig:rdt}
\end{figure}

In Fig.~\ref{fig:rdt}, we show an example of the the behaviour of the RDT in the case where $\epsilon=5$.  We notice how the increasing trend of the receiving rate ceases for probing rates greater than 7 Mbps.  As as indication, we display a metric related to ours: the available bandwidth as estimated by Pathload.

\subsection{Graphical Model}
The available bandwidth of each path is determined by all the links it is composed of, more precisely, by the minimum of all these links' available bandwidths. The joint probability distribution representing the functional relationship between links and paths of a topology is very complex.  To  capture this relationship, we use factor graphs, which allow us to calculate marginals for each path and link more simply.

We construct the graphical model using variable and factor nodes.  Each link $\ell$ and path $p$ have a variable node in the factor graph ($x_{\ell}$ and $y_p$: the probability that the probabilistic available bandwidth is $r$), which are interconnected through $\min$ factor nodes $f_{y_p}$.  Each link variable node is also connected to a factor node $f_{x_{\ell}}$ that represent prior knowledge about the associated link variable.  For each measurement we make, we add a variable node $z_k$ equal to the binary outcome of measurement $k$ and connect it the the variable node of the measured path through a factor node $f_z = L(z|y_p)$, representing the likelihood function.  After each measurement, we run the belief propagation algorithm; a procedure that updates the estimated marginal distributions of all variables in the graph based on the newly acquired information.  When the algorithm terminates, we can compute marginals for any link and path in the graph.

\begin{figure}[!h]
	\centering
	\subfigure[$Pr(z_{RDT} = 1)$ as a function of the difference between the probing rate and estimated available bandwidth.  Each data point is obtained by averaging ten RDT with $\epsilon=5$ over five different paths.  The best fit is obtained by performing a regression for parameters $\alpha$ and $y_p$.]{\label{fig:nm_global}\includegraphics[width=0.7\linewidth]{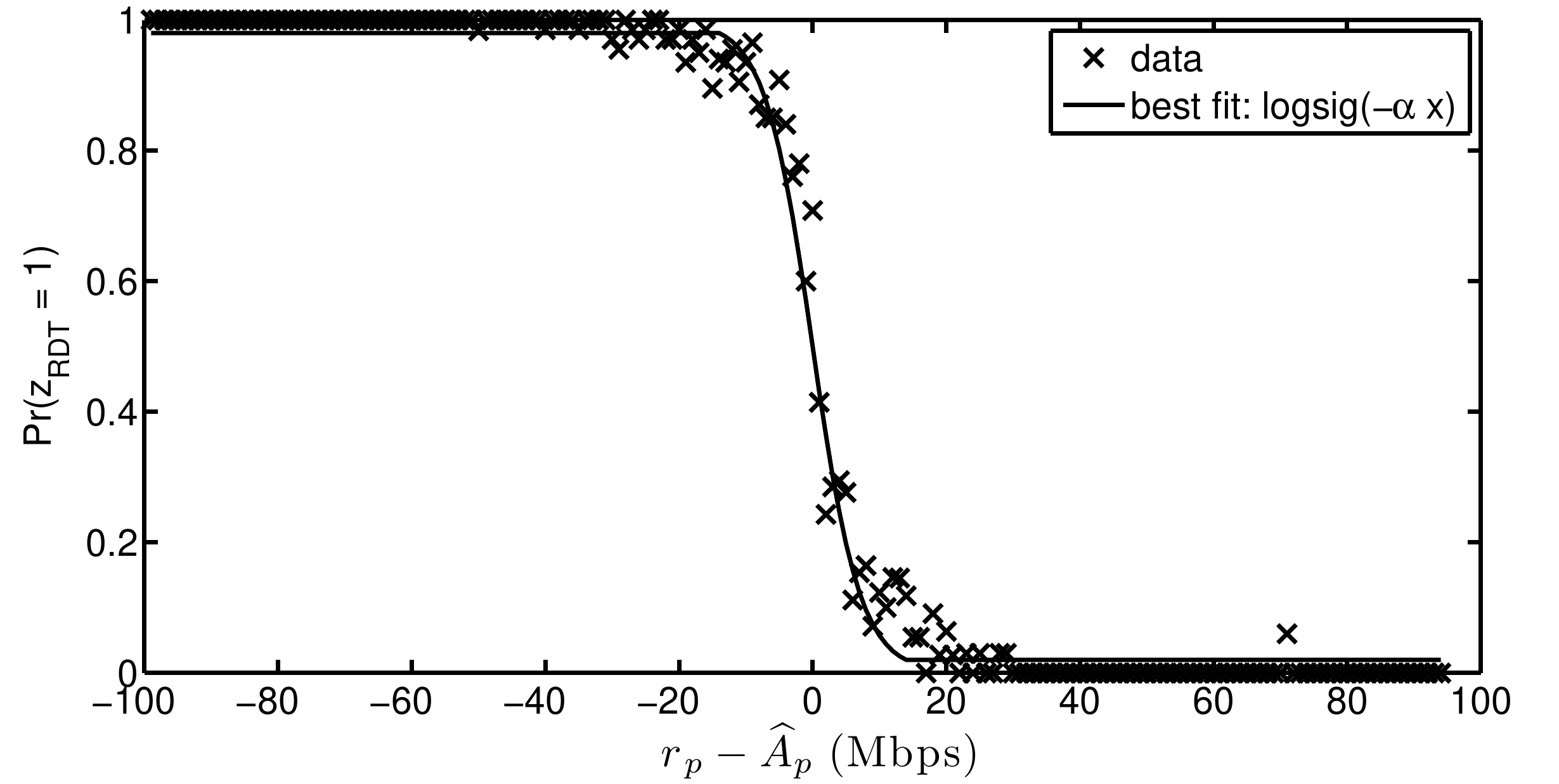}}\\
	\subfigure[Probability of $z_{RDT} = 1$ as a function of the probing rate $r_p$ for four different paths.  The best fit are identical to the one in (a).\label{fig:nm_ind}]{\includegraphics[width=0.7\linewidth]{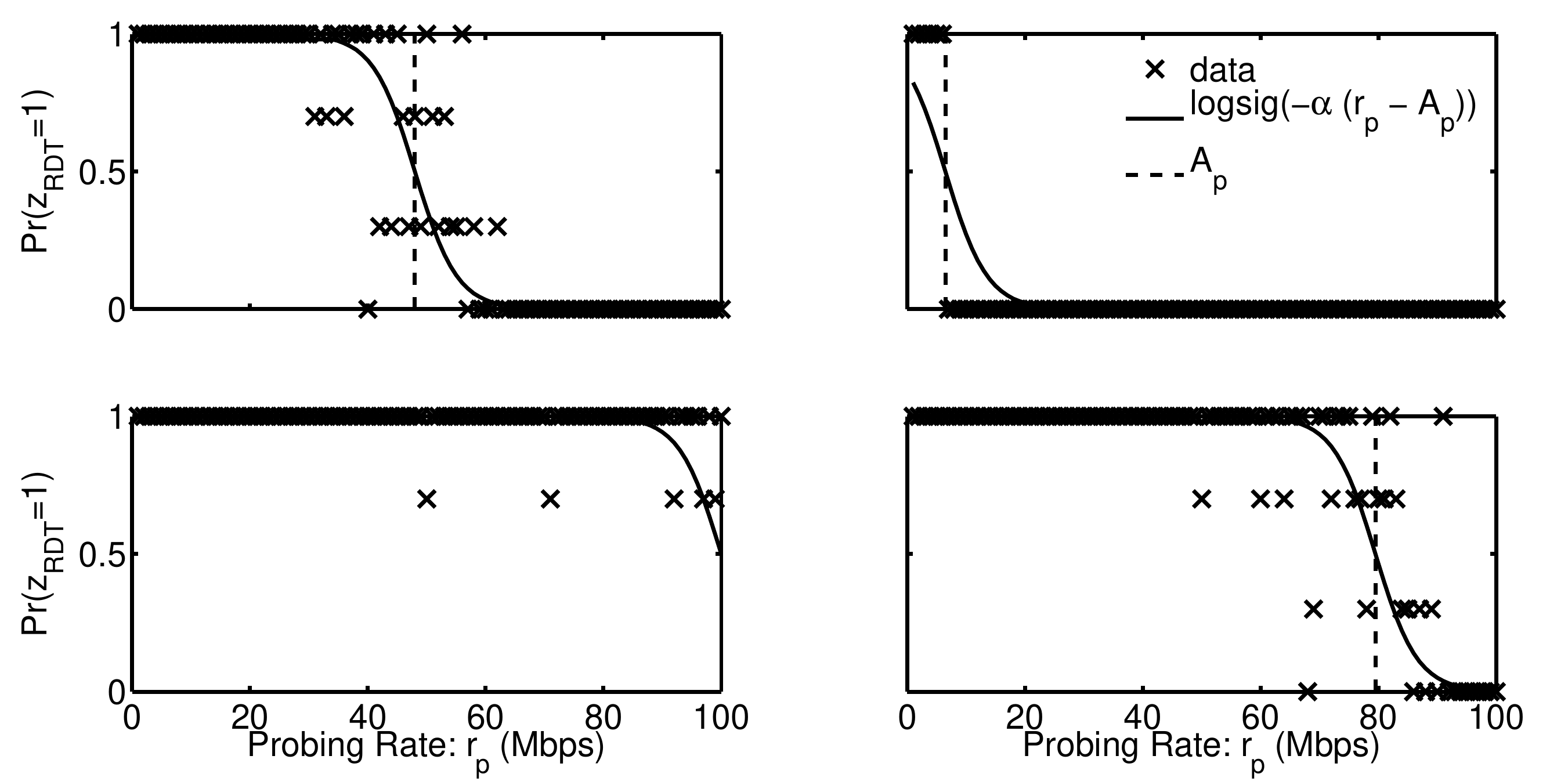}}
	\caption{Empirical data and regression fit for likelihood model.}
	\label{fig:likelihood}
\end{figure}

Intuitively, when the probing rate $r_p$ is well below $y_p$, we expect the probability of observing $z=1$ to be very high and, similarly, when $r_p$ is well over $y_p$, this probability should be very close to zero.  However, when $r_p$ is equal to or very close to $y_p$, we expect this probability to be closer to 0.5.  Although a simple step function looks like a good match, it is too aggressive as we expect higher levels of noise when we probe around $y_p$.  Based on these intuitive expectations and experimental data (Fig.~\ref{fig:likelihood}), we adopt the likelihood model $L(z=1|y_p,r)={\mathrm{logsig}}(-\alpha(r-y_p))$ for the measurements, where $\alpha$ is a small positive constant learned empirically.  However, to determine the value of $\alpha$ we first need to estimate $y_p$.  We decide to co-jointly estimate the values of $y_p$ along with the constant $\alpha$ through a single regression procedure where we determine the best fit by minimizing the MSE. 

The sigmoid function rapidly decays to zero when the probing rate is greater than the available bandwidth, even for the best possible parameter fit.  We wish to be careful and prefer a slightly less aggressive approach where we assign some likelihood to unexpected measurements outcomes at all ingress rates.  For that reason, we introduce a small constant $\kappa$ and bound our likelihood function to lie in the range $[\kappa ,1-\kappa ]$.  

We note that our estimation procedure is not sensitive to the exact choice of $\alpha$, which specifies the rate of decay of the sigmoid function. Moreover, in experiments conducted on different topologies, days, and times-of-day, we have observed that the estimated $\alpha$ values occupy a small range of values. The values are related to the variability of the path available bandwidths over the measurement interval. These observations suggest that it is possible to execute the training procedure rarely, or to completely forego it and choose a conservative value. The latter strategy does not bias the estimation procedure, but it does slow the rate of convergence.

\begin{figure}[!h]
\centering
\subfigure[Logical topology]{
		\begin{tikzpicture}[auto,inner sep=1mm]
			\node (router 4) at (0,0) [router] {4};
			\node (router 3) at (0,1) [router] {3};
			\node (router 1) at (-1,2) [router] {1};
			\node (router 2) at (1,2) [router] {2};
			\node (router 5) at (-1,-1) [router] {5};
			\node (router 6) at (1,-1) [router] {6};	
			
			\draw (router 1) to node [swap] {$\ell_1$} (router 3) ;
			\draw (router 2) to node {$\ell_2$} (router 3) ;
			\draw (router 3) to node {$\ell_3$} (router 4) ;
			\draw (router 4) to node [swap] {$\ell_4$} (router 5);
			\draw (router 4) to node {$\ell_5$} (router 6);
		\end{tikzpicture}
		\label{fig:logicaltopology}
}
\subfigure[Factor graph representation with measurement node $f_z$ attached to path $p_1$.]{
\begin{tikzpicture}
			\node[varnode] (x1) {$x_{\ell_1}$};
			\node[varnode] (x2) [right of=x1] {$x_{\ell_2}$};
			\node[varnode] (x3) [right of=x2] {$x_{\ell_3}$};
			\node[varnode] (x4) [right of=x3] {$x_{\ell_4}$};
			\node[varnode] (x5) [right of=x4] {$x_{\ell_5}$};
			
			\node[funcnode] (fx1) [above of=x1] {$f_{x_{\ell_1}}$};
			\node[funcnode] (fx2) [above of=x2] {$f_{x_{\ell_2}}$};
			\node[funcnode] (fx3) [above of=x3] {$f_{x_{\ell_3}}$};
			\node[funcnode] (fx4) [above of=x4] {$f_{x_{\ell_4}}$};
			\node[funcnode] (fx5) [above of=x5] {$f_{x_{\ell_5}}$};	
			
			\draw (x1) -- (fx1);
			\draw (x2) -- (fx2);
			\draw (x3) -- (fx3);
			\draw (x4) -- (fx4);
			\draw (x5) -- (fx5);
			
			\node[funcnode] (fy1) [below of=x1,xshift=5mm,yshift=-10mm] {$f_{y_{p_1}}$};
			\node[funcnode] (fy2) [below of=x2,xshift=5mm,yshift=-10mm] {$f_{y_{p_2}}$};
			\node[funcnode] (fy3) [below of=x3,xshift=5mm,yshift=-10mm] {$f_{y_{p_3}}$};
			\node[funcnode] (fy4) [below of=x4,xshift=5mm,yshift=-10mm] {$f_{y_{p_4}}$};
			
			\node[varnode] (y1) [below of=fy1] {$y_{p_1}$};
			\node[varnode] (y2) [below of=fy2] {$y_{p_2}$};
			\node[varnode] (y3) [below of=fy3] {$y_{p_3}$};
			\node[varnode] (y4) [below of=fy4] {$y_{p_4}$};
			
			\draw (y1) -- (fy1);
			\draw (y2) -- (fy2);
			\draw (y3) -- (fy3);
			\draw (y4) -- (fy4);
			
			\draw (fy1.north) -- (x1.south);
			\draw (fy1.north) -- (x3.south);
			\draw (fy1.north) -- (x4.south);
			
			\draw (fy2.north) -- (x1.south);
			\draw (fy2.north) -- (x3.south);
			\draw (fy2.north) -- (x5.south);
			
			\draw (fy3.north) -- (x2.south);
			\draw (fy3.north) -- (x3.south);
			\draw (fy3.north) -- (x4.south);
			
			\draw (fy4.north) -- (x2.south);
			\draw (fy4.north) -- (x3.south);
			\draw (fy4.north) -- (x5.south);
			
			\node[funcnode] (fz) [below of=y1] {$f_z$};
			
			\draw (y1) -- (fz);	
			
			\node[varnode] (z) [below of=fz] {$z$};	
			
			\draw (z) -- (fz);
		\end{tikzpicture}		
\label{fig:factor_graph}
}
\caption{Six nodes topology with $N=5$ links ($\ell_1,...,\ell_5$) and $M=4$ paths ($p_1,...,p_4$): 1-3-4-5, 1-3-4-6, 2-3-4-5, 2-3-4-6.}
\label{fig:topo_factorgraph}
\end{figure}
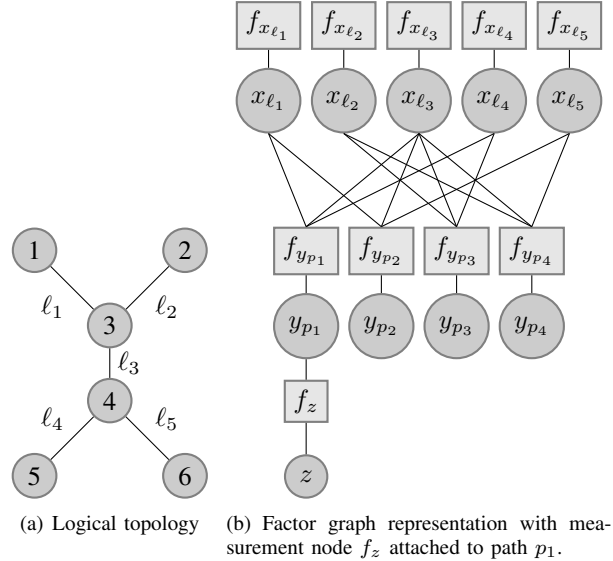

\noindent {\bf Example:} In Figure~\ref{fig:logicaltopology}, we show an example of a simple logical topology of a network, which is required to determine how path and link variable nodes are connected in the factor graph.  In this example, there are six nodes interconnected using $N=5$ different links labeled $\ell_1,...,\ell_5$ and we consider $M=4$ paths ($p_1,...,p_4$) where nodes 1 and 2 are the sources and nodes 5 and 6 are the destinations. In Figure~\ref{fig:factor_graph}, we can see the factor graph representation of this logical topology after one measurement has been made of $p_1$.

We then construct a likelihood model for the network we used for our experiments using $\epsilon = 5$, $\delta = 0.5$ and a range of values where $B_{min} = 1$ and $B_{max} = 100$.  We first gather data from five different paths: 500 measurements from non-consecutive packet trains at each rate between $B_{min}$ and $B_{max}$.  We then repeat this experiment five times at different periods of the day resulting in 25 sets of 500 measurements.  In Fig.~\ref{fig:nm_ind}, we depict $Pr(z_{RDT} = 1)$ as a function of $r_p$ for four different paths (results did not vary with time of the day).  We normalize each of the 25 experiments and combine all the data in a single plot as a function of $r_p - \widehat{y}_p$. The result is shown in Fig.~\ref{fig:nm_global} where each data point is the result of averaging all values which had the same value of $r_p - y_p$; all experiments for which the distance between $r_p$ and $y_p$ is identical.  

\subsection{Active Path Selection}
\label{ssec:path}

A conventional approach to sample many paths is using non-adaptive uniform techniques~\cite{cas:07}: the path could be chosen either randomly or using an approach like round robin (RR).  However, since making each measurement is costly in terms of time and bandwidth, there is an incentive to reduce the total number of measurements.  By actively selecting the path and the rate using information collected up to that point, adaptive sampling can provide significant gains in terms of time and resources.  The drawback of this approach is that it is difficult to estimate the informativeness of each potential measurement.  

We now describe two greedy active learning procedures to select the path to probe at each iteration. Both algorithms are probabilistic in nature: they determine the probability that each path is chosen (denoted $f_{probe}^k(p)$, and then the choice is accomplished by making a random selection according to the specified probabilities.
The first algorithm is called weighted entropy (WE).  Paths with high entropy have more uncertainty and therefore there is more information to gain by measuring those paths. 
With WE, paths with high entropy are assigned a larger probability of being selected (the probability of probing a path that already satisfies our stopping criteria is zero).
We assign a probing probability using the following relation: 

$$ f_{probe}^k(p) = \left\{\begin{array}{ll}-\sum_{r=B_{min}}^{B_{max}} \hat{p}(y_p = r) \cdot \log \hat{p}(y_p = r) & \beta_p > \beta\\ 0 & \beta_p \leq \beta \end{array}\right.$$

\noindent 
We then normalize to ensure $\textstyle{\sum}_{p \in \mathcal{P}} f_{probe}^k = 1$.

The second algorithm, called weighted confidence interval (WCI), bases path selection on the current confidence range $\beta_p$ of each path.  A smaller confidence range signifies that most of the distribution is localized in a given region, which indicates a higher level of confidence in the estimate.  On the other hand, a large $\beta_i$ signifies that the distribution is spread out over a greater range of potential values, so it is harder to produce an accurate estimate.  Also, since we are using $\beta_p$ as a stopping criterion, it encourages the algorithm to terminate faster.  The path probing probability function for WCI is
$$ f_{probe}^k(p) = \left\{\begin{array}{ll} \beta_p & \ \beta_p > \beta\\ 0 &  \beta_p \leq \beta \end{array}\right.$$
and we again normalize to ensure the probabilities sum to one.

\subsection{Active Probing Rate Selection}
\label{ssec:active_rate}

Previous approaches bounds to determine the interval within which the available bandwidth lies.  To decide on the probing rate, they use either a linear or a binary search approach and choose the mean of the bounds.  We adopt a different approach and choose the rate that bisects the distribution: the median of the posterior distribution of the path.  By probing at the median, there is equal probability (according to our current knowledge) that the available bandwidth is smaller or greater than the probing rate.  We therefore maximize the expected information gain from our measurement; it is equivalent to conducting a probabilistic binary search for the available bandwidth on path $p$~\cite{cas:07}.  By using a probabilistic rather than deterministic approach in rate selection, hard decisions (which could be incorrect) are not enforced.

\section{Simulation and Experimental Results}
\label{sec:exp}

\subsection{Simulation Results}
\label{ssec:sim}

The purpose of the simulations is to explore the efficacy of our proposed learning strategies. These are not network simulations, so they do not test modelling assumptions at all (that is the purpose of the PlanetLab experiments).  We use the HOT topology generated using Orbis to run our simulations~\footnote{http://www.sysnet.ucsd.edu/~pmahadevan/topo\_research/topo.html}.  It includes 939 nodes and 988 links, from which we extract subsets of $M=50,100,150,200,250$ paths.  We assign link capacities using a uniform distribution between $[1,100]$ and generate a total of 500 topologies.  At each iteration, probe outcomes are generated according to our likelihood model. As stopping criteria, we use $\eta = 0.95$, $\beta = 10$ and a maximum of 10000 measurements.  We compare three path selection algorithms (RR, WE and WCI) and also show the average number of measurements and accuracy required when our active learning algorithm is run independently and sequentially on each path (SEQ).

\begin{figure}[!h]
	\centering
	\includegraphics[width=0.7\linewidth]{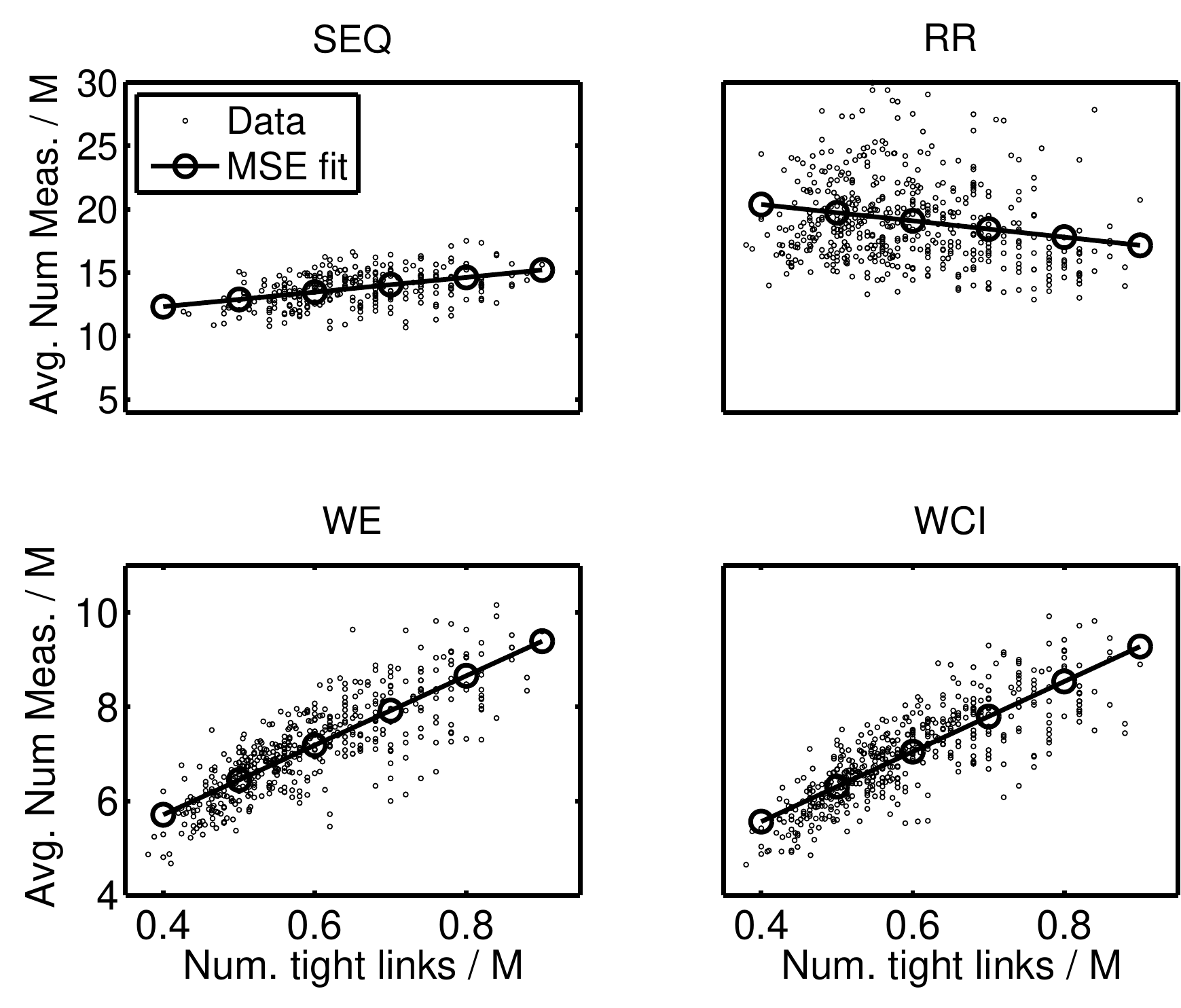}
	\caption{Simulated average number of measurements as a function of the number of tight links in the topology.  Both values are normalized by the number of paths $M$.  We show all the simulated values and a first degree polynomial fit for each technique.}
	\label{fig:simulation}
\end{figure}

An estimate is considered accurate if the real available capacity lies within the confidence interval: $lb_p \leq y_{p} \leq ub_p$.  The average accuracy of all methods is satisfactory; it is always higher than the confidence level $\eta=0.95$.  However, it is in terms of average number of measurements that the adaptive techniques (WE, WCI) stand out (see Fig.~\ref{fig:simulation}).  The average number of measurements required by WCI is between $46-73\%$ lower than the number required by RR and $39-55\%$ lower than SEQ.  Employing the factor graph in a naive manner can be disadvantageous because noisy measurement outcomes can spread uncertainty to other path estimates (see RR).  WE and WCI provide important savings in terms of time and measurements without affecting the accuracy, but since WCI is slightly better in terms of average number of measurements, we will use WCI for our online experimentations.

\subsection{Obtaining and Calculating $r_p$ and $r'_p$}

For our online experiments, we have deployed our measurement software coded in C on various nodes on the PlanetLab network\footnote{http://www.planet-lab.org/}.  Although this network was once believed to be too heavily loaded, Spring et al.\ explained that PlanetLab has evolved and this is no longer true~\cite{spr:06}.  The authors also describe how other myths, such as load prevents accurate latency measurements and load prevents sending precise packet trains, can be avoided with best practices.  To gather measurement data we have implemented software able to transmit and receive sequences of packet trains.  A single measurement consists of sending $N_t$ sequences (trains) of $L_s$ UDP packets of $P_{size}$ bytes at a rate $r_p$ and observing the rate $r'_p$ at the receiver side.  We then take the median of $r'_p$ obtained from each of the $N_t$ trains and perform the RDT to obtain $z$.

Typically, the receiving rate is calculated by dividing the total number of bytes received by the amount of time that elapsed between the reception of the first and last packet.  However, during our measurements, we have observed delays on the sender side (probably due to task interruption) between the departure of two consecutive packets ($t_i > t_{i-1} + \tau$).  These delays would bias the receiving rate if we were to calculate it using this classical method.  Therefore, to calculate the rate $r'_p$ at the receiver side, we use inter-arrival time between valid packets.  Upon reception of the last packet of a train, we construct a set $V$ of all the indices $t>1$ of valid packets. A packet is labelled as invalid if the difference between its departure and the departure of the previous packet is greater than $\tau$.  The equation we use to calculate $r'_p$ is the following:

\begin{equation}\label{eq:recrate}
r'_p = \frac{\sum_{t \in V} t_r(t) - t_r(t-1)}{|V| \cdot P_{size}},
\end{equation}

\subsection{Experimental Results}

\begin{figure}[!h]
	\centering
\includegraphics[width=0.7\linewidth]{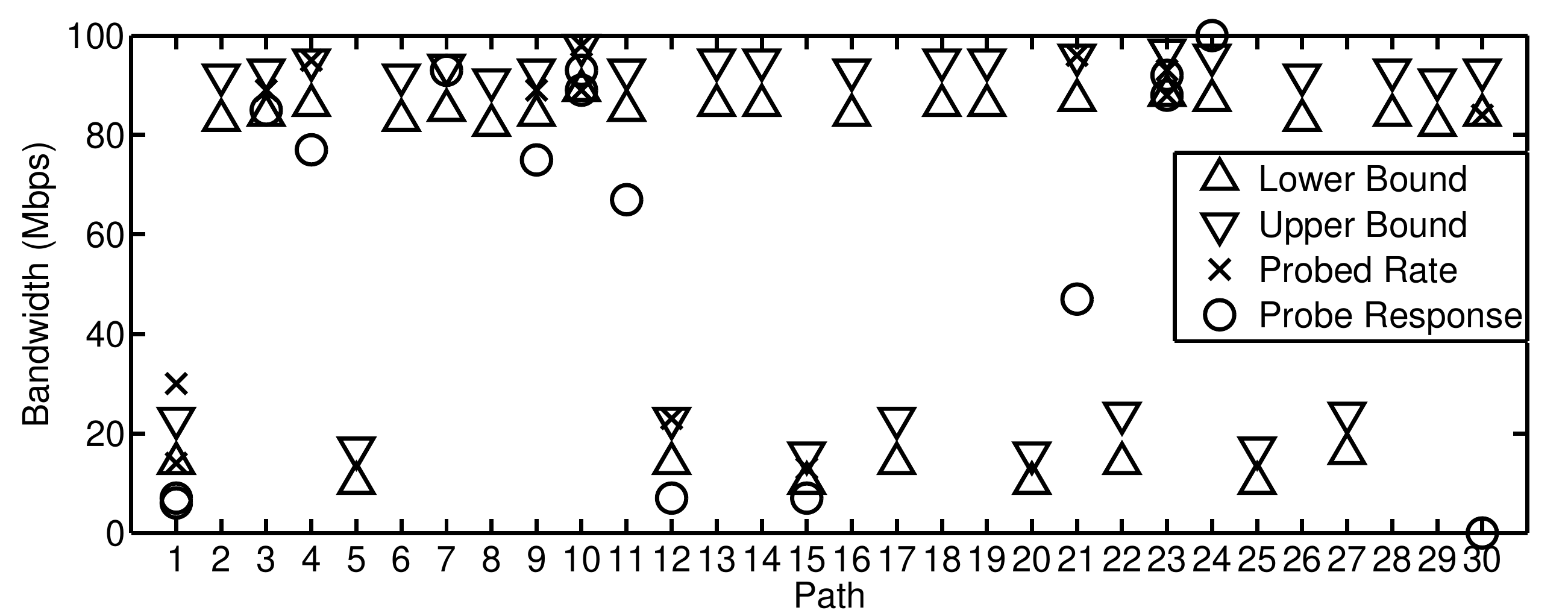}
	\caption{Bounds of the confidence intervals for a 30 paths topology.	\label{fig:pl_exp}}
\end{figure}

We ran an online experiment using a different topology with $M=30$ paths\footnote{All possible end-to-end paths between the following six PlanetLab nodes: planetlab3.csail.mit.edu, planetlab-1.cs.unibas.ch, planlab1.cs.caltech.edu, planetlab2.acis.ufl.edu, planetlab1.cs.stevens-tech.edu, planetlab2.csg.uzh.ch.} and $N=65$ logical links with the following parameters: $\beta = 10$ Mbps, $\eta = 0.95$, $B_{min} = 1$ Mbps, $B_{max} = 100$ Mbps, $\epsilon = 5$ and the path selection algorithm is WCI.  We show the lower and upper bounds obtained for all $M=30$ paths during one experiment in Fig.~\ref{fig:pl_exp}.  We also display the outcome of transmission tests conducted at the end of the estimation interval. We performed four tests on four disjoint paths (for a total of 16 tests per run) by sending trains of 2400 packets of 1000 bytes (the equivalent of 60 seconds of video encoded at 320 kbps) and observing the output rate. In each of the tests, the sending rate of the train is different --- the lower bound of the confidence interval $lb$, the lower bound plus $5$ Mbps, the upper bound of the confidence interval $ub$, and $5$ Mbps above the upper bound. 

\begin{figure}[!h]
	\centering
	\subfigure[Empirical probability of observing $z=1$ averaged over 17966 measurements.]{\label{fig:raw_trainsize}\includegraphics[width=0.5\linewidth]{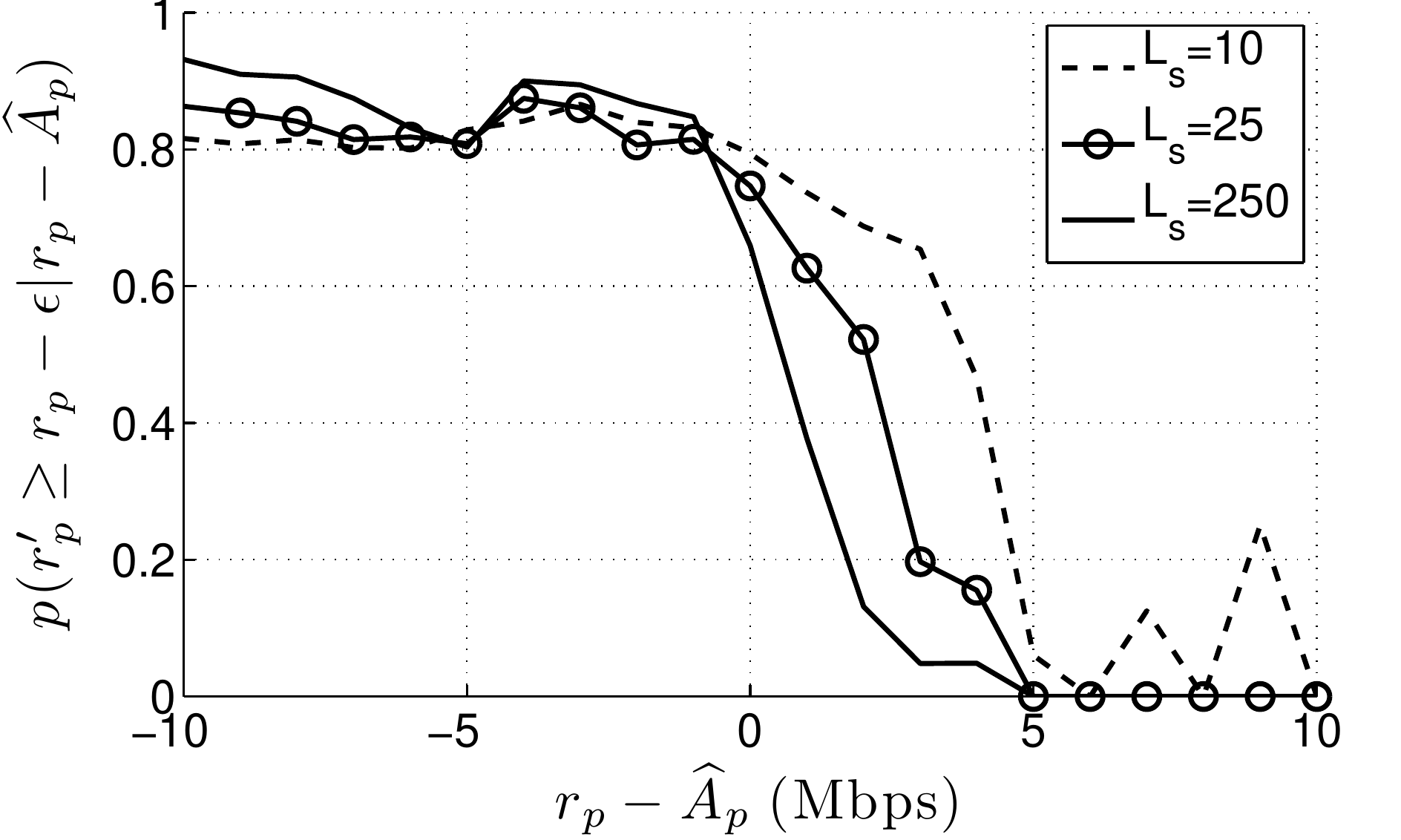}}\\
	\subfigure[Average number of bytes used per path during the estimation procedure.]{\label{fig:bytes_per_trainsize}\includegraphics[width=0.5\linewidth]{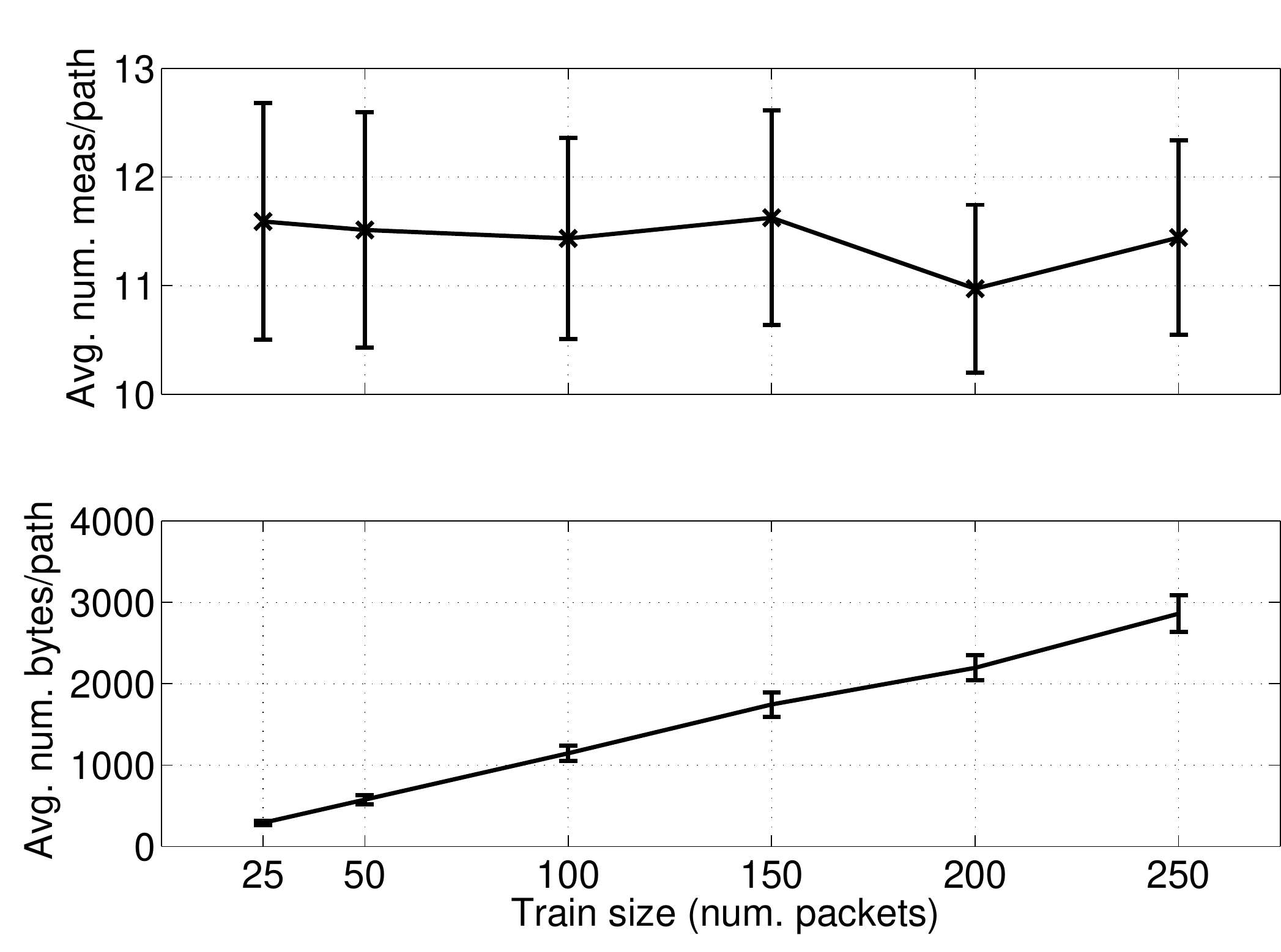}}\\
	\subfigure[Accuracy of each transmission test (different probing rates) averaged over 80 tests.]{\label{fig:live_accuracy_train_size}\includegraphics[width=0.5\linewidth]{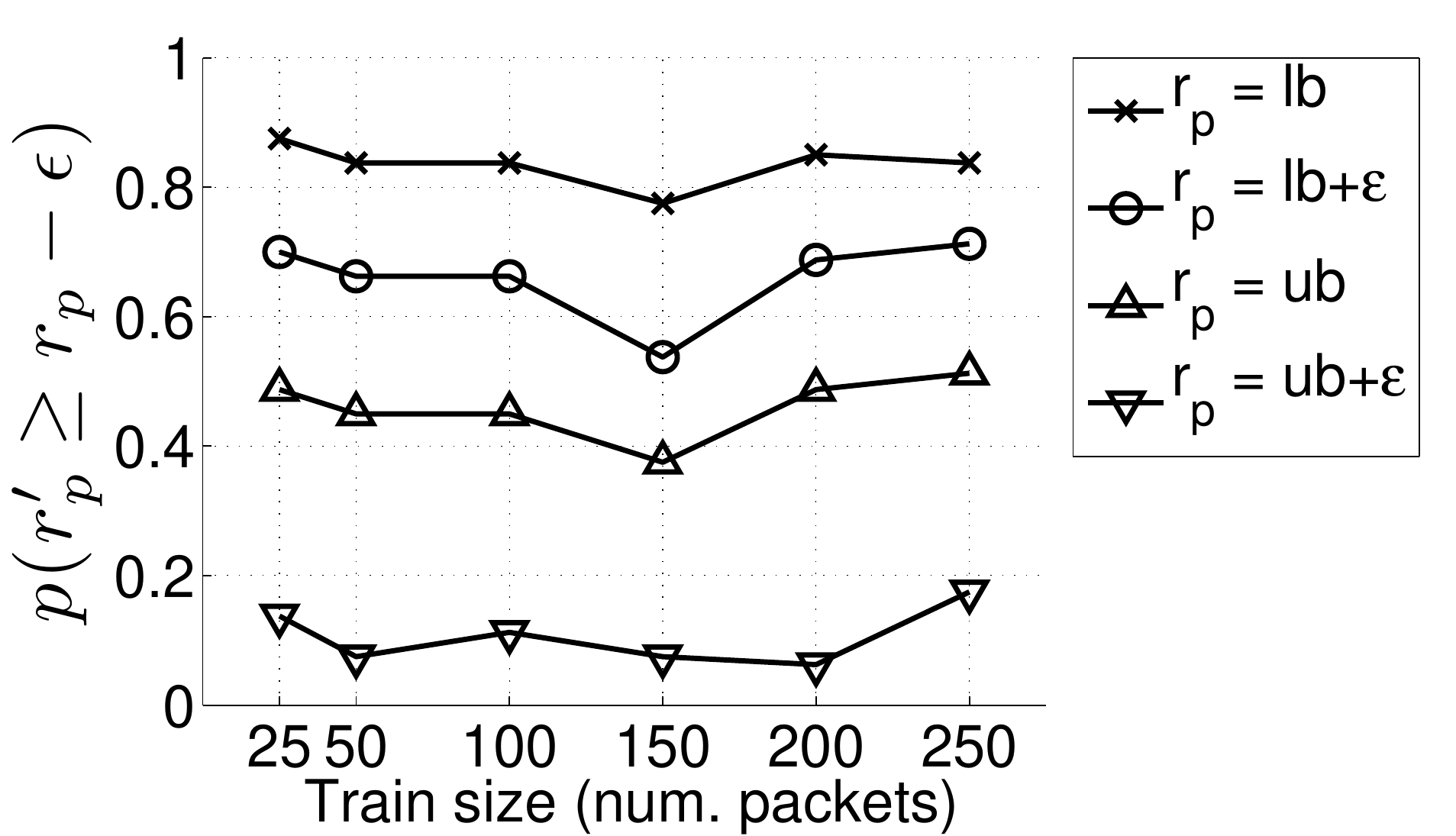}}
	\caption{Impact of train size on measurement outcome $z$ and accuracy for different train sizes.}
\end{figure}

An important measurement parameter is the number of packets in each probe train; clearly it is desirable to reduce this number to limit the measurement load, but we anticipate that more rate estimation errors will occur as the probe length is reduced. In Fig.~\ref{fig:raw_trainsize}, we study the impact of the number of packets in a train $L_s$ on the measurement outcome $z$ by collecting traces with 250 packets and then using only a fraction of them to evaluate the egress rate and calculate $z$.  We cluster all 17966 measurements (gathered from 6 nodes during 20 measurement sequences) according to the difference between the probing rate and our estimate of the available bandwidth (here we use the maximum a posteriori (MAP) estimate). Using this data we calculate the empirical probability of observing $z=1$. In the noiseless case, assuming perfect estimates of the available bandwidths, this would look like a step function centred at 0.  When the probing rate is smaller than the available bandwidth, the incertitude does not vary greatly with the train size $L_s$.  The main difference occurs when the probing rate is greater than the available bandwidth; as we decrease $L_s$, the incertitude decays much more slowly and in the case of $L_s = 10$, there is even a $20\%$ chance of obtaining $z=1$ when probing 10 Mbps above the estimated available bandwidth.  With the exception of this slower decay phase, using 25 and 250 packets for the measurements results in similar behaviour.  In Fig.~\ref{fig:bytes_per_trainsize}, we show the average (over twenty experiments with two different same size topologies) number of measurements and bytes per path required to complete the estimation procedure as a function of the train size.  Since the number of measurements is constant for all values of $L_s$, we observe a linear growth in the number of bytes required to achieve the desired accuracy.  

To evaluate the predictive properties of our algorithm, we calculate the empirical probability (from a total of 80 tests) that the egress rate exceeds the ingress rate less the tolerance factor $\epsilon$.  The empirical probability, shown in Fig.~\ref{fig:live_accuracy_train_size}, when probing at the upper bound of the confidence interval is close to $0.5$, the target $\delta$ for our experiments.  This value drops to around $0.1-0.2$ when exceeding the upper bound.  Probing at the lower bound leads to an empirical probability of $0.8-0.9$. From these results, we can conclude that from the estimated confidence intervals, we can determine the probability of avoiding congestion for various rates.  The fact that performance varies very little with the number of packets in the train indicates that, for this network at least, 25 packets per train would suffice while providing significant savings in terms of bytes used.  The results displayed in this figure, along with Fig.~\ref{fig:raw_trainsize}, indicate that our technique slightly underestimates the probabilistic available bandwidths.

\begin{table}[!h]
	\caption{Average time and bytes used by Pathload and WCI for $M=30$ paths topology over 5 runs.\label{tab:pathload}}
	\centering
	\begin{tabular}{c|cc}
		& seconds / path & kbytes / path \\
		\hline
		Pathload & $27.0 \pm 0.8$ & $10806 \pm 1058$\\
		WCI & $7.1 \pm 0.3$ & $612 \pm 18$
	\end{tabular}
\end{table}

Although probabilistic available bandwidth is a different metric than classical available bandwidth, the two metrics are related.  It is interesting to compare our estimation methodology with Pathload~\cite{jai:03}, both to assess measurement overhead and to examine the extent of correlation between the two metrics.. Using the same topology described above ($M=30$, $N=65$), we run both Pathload sequentially on every single path and our algorithm (WCI and 25 probes/train) and compare our confidence interval to Pathload's variation range, number of bytes transmitted and time elapsed.  The results are summarized in Table~\ref{tab:pathload}.  We can see that WCI provides significant gains in terms of measurement latency ($75\%$ reduction in seconds) and overhead ($95\%$ reduction in bytes).  In terms of confidence intervals, they overlapped for $53\%$ of the paths - $76\%$ if we tolerate a 2Mbps error.  Since the metrics are different, a complete correspondence between the estimates is not expected. But despite the fact that they focus on different metrics, both estimation techniques strive to examine the same path property (at what rate can probe trains be sent without inducing congestion). The experiments highlight the correspondence between the estimates and indicate how dramatic the time and measurement overhead savings can be.

\section{Related Work}
\label{ssec:related_work}

Our proposed algorithm is designed to estimate the probabilistic available bandwidth of multiple paths, and it achieves efficiency by sharing information through a probabilistic graphical model and using active learning to choose the most informative measurements. The metric we estimate is fundamentally different than the traditional available bandwidth metric, defined in terms of unused capacity, but there are strong connections. We now briefly review the major methods that have been proposed for available bandwidth estimation. Almost all of these focus on estimating available bandwidth for a single path; the exception is BRoute, which addresses multiple paths, as discussed below.

Many approaches and software tools aimed at estimating available bandwidth of end-to-end network paths using active measurements have been proposed over the past few years: Spruce~\cite{str:03}, IGI/PTR~\cite{hu:03}, pathChirp~\cite{rib:03}, Pathload~\cite{jai:03}, abget~\cite{ant:06} (see~\cite{jai:07} for a thorough survey and description of each technique). In self-congestion probing, trains of packets are sent at a known rate on the path of interest and if the rate is greater than the available bandwidth of one or more of the links, then extra arriving traffic will accumulate at the links' buffer causing an increasing trend in the one-way delays of consecutive packets and an egress rate at the receiver lower than the ingress rate.  On the other hand, if the probing rate is lower than the available bandwidth, there will be no noticeable trend in the inter-packet spacing and the egress rate should be approximately equal to the ingress rate.  Then, to estimate the available bandwidth, the strategy is to identify where the transition between these two behaviours occurs. In techniques like pathChirp~\cite{rib:03} and Pathload~\cite{jai:03} the sender varies the probing rate iteratively until it is possible to converge to the available bandwidth of the path.  To account for noise and the variability of the available bandwidth, Jain and Dovrolis estimate a variation range for the available bandwidth (instead of a single estimate) based on one-way delays using Pathload~\cite{jai:03}.  Pathload and pathChirp employ deterministic inference strategies, introducing relatively ad-hoc outlier elimination rules and using averaging to robustify the statistics derived from their measurements. 

Liu et al.\ study in more depth the transition point and the relation between the input/output ratio and the input rate~\cite{liu:08}.  They derive a stochastic response curve that is tightly lower bounded by its fluid counterpart (used by previously referenced techniques) and conclude that this transition point is not exact.  The idea of stochastic service curves, which expresses the service given to a flow by the network in terms of a probabilistic bound, was also used by Ciucu et al.\ to provide bounds on end-to-end delay~\cite{ciu:05}.  In the context of available bandwidth estimation, Liebeherr et al.\ propose an elegant theoretical framework based on min-plus algebra, but only for worst-case deterministic estimations~\cite{lie:07}.

Other approaches have been proposed previously to perform network-wide estimations without overloading the network and/or consuming a large amount of resources.  Song and Yalagandula proposed $S^3$ to measure real-time end-to-end network properties such as latency and loss rate~\cite{son:07}.  They  measure only a subset of the network, by choosing paths according to the observed load on the links and at the end nodes, to infer statistics about the entire network.  However, their work does not address the problem of estimating available bandwidth.  In~\cite{hu:05}, Hu and Steenkiste introduce BRoute; a scalable available bandwidth estimation system based on route sharing.  They assume that most Internet bottlenecks are on path edges and use the fact that links near end-nodes are often shared by many paths.  Although some of the intuition behind BRoute and our work is similar, the core implementation differs; it is not based on a probabilistic framework.

\section{Conclusion and Future Work}
\label{sec:conc}

In this paper, we defined probabilistic available bandwidth, a quantity different from standard available bandwidth, as the highest rate at which data can be sent such that the egress rate is equal to the ingress rate (within a level of tolerance) with probability $\delta$.  The methodology we present addresses estimation on multiple paths and is the first to employ probabilistic inference, graphical models, and active learning. We have implemented measurement software and deployed it on the PlanetLab network.  We used our software to identify an appropriate likelihood model.  We designed a procedure to map a topology to a factor graph format and used a loopy belief propagation algorithm to update estimates of the marginals of the available bandwidths of the paths.  We took advantage of the statistics derived from these marginals to select the most informative path and rate at each iteration based on active learning techniques.  From our simulations, our active learning approach dramatically reduces the number of measurements (by more than $50\%$) required to achieve the same level of accuracy as passive approaches such as round robin.  We ran online experiments on the PlanetLab network and observed that very few probes are required to perform our estimation procedure accurately.

\subsection{Future Work and Challenges}

Although these initial results are very encouraging and promising, there are many avenues to explore to improve our method.  We need to explore in more depth the trade-off between accuracy and number of measurements and take advantage of the fact that we can achieve the desired level of accuracy with relatively few packets per train.  By gaining a better understanding of this, we could formulate the problem in terms of optimizing a resource (time, bytes) budget while achieving a required accuracy level.  Also, we need to develop an automatic procedure to learn what is the appropriate number of packets in a train for a given network (for some networks we might need long trains, for others short trains might suffice).  We are currently investigating an approach that employs chirps in order to scan a wider range of bandwidths with a single measurement~\cite{rib:03}.  To further reduce the number of measurements required, we wish to use a more informative prior for the links than the uniform distributions we are currently using.  By adopting priors that encourage sparse solutions (a small number of tight links), we hope to accelerate our algorithm.  Finally, we would also like to deploy our software tool on a different platform than PlanetLab and on a larger number of nodes.  We will explore other testing procedures to validate the estimation obtained with our tool.

\bibliographystyle{IEEEtran}
\bibliography{enabe}

\end{document}